\documentclass[11pt]{article}

\input psfig.sty

\usepackage{amssymb}

\def\openone{\leavevmode\hbox{\small1\kern-3.8pt\normalsize1}}%

\def\mf{{\mbox{\tiny\em MFA}}}

\def\ba{\begin{array}}
\def\ea{\end{array}}
\def\be{\begin{equation}}
\def\ee{\end{equation}}
\def\bea{\begin{eqnarray}}
\def\eea{\end{eqnarray}}

\renewcommand{\thefootnote}{$\dagger$}

\begin{document}
\thispagestyle{empty}
\begin{center}
\vspace*{-1cm}
{\Large \bf Two flavor color superconductivity in nonlocal chiral
quark models}
\vspace*{.8cm} \\
{ R. S. Duhau$^a$, A. G. Grunfeld$^a$ and N.N. Scoccola$^{a,b,c}$}
\vspace*{0.4cm} \\
$^a$ {\em Physics Department, Comisi\'on Nacional de Energ\'{\i}a At\'omica, \\
 Av.Libertador 8250, (1429) Buenos Aires, Argentina.}  \\
$^b$ {\em CONICET, Rivadavia 1917, (1033) , Argentina.}\\
$^c$ {\em Universidad Favaloro, Sol{\'\i}s 453, (1078) Buenos Aires,
Argentina}
\vspace*{.5cm}
\date{\today}
\begin{abstract}
We study the competence between chiral symmetry restoration and
two flavor color superconductivity (2SC) using a relativistic
quark model with covariant nonlocal interactions. We consider two
different nonlocal regulators: a Gaussian regulator and a
Lorentzian regulator. We find that although the phase diagrams are
qualitative similar to those obtained using models with local
interactions, in our case the superconducting gaps at medium
values of the chemical potential are larger. Consequently, we
obtain that in that region the critical temperatures for the
disappearance of the 2SC phase might be of the order of 100-120
MeV. We also find that for ratios of the quark-quark and
quark-antiquark couplings somewhat above the standard value 3/4,
the end point and triple point in the $T-\mu$ phase diagram meet
and a phase where both the chiral and diquark condensates are
non-negligible appears.
\end{abstract}
\vspace*{.3cm}
\end{center}

\vspace{1cm}

\renewcommand{\thefootnote}{\arabic{footnote}}
\setcounter{footnote}{0}

\section{Introduction}

The understanding of the QCD phase diagram has become one of the
most interesting issues in the physics of strong interactions.
Such phase diagram is relevant to phenomena in the early universe,
in the interior of neutron stars and in relativistic heavy ion
collisions. Already in the seventies it was suggested that there
should be two distinct phases: a low temperature and density phase
in which quarks and gluons are confined within hadrons and chiral
symmetry is broken, and a high temperature and density phase (the
so-called quark-gluon plasma) in which these particles are
deconfined and chiral symmetry is restored. Although the possible
existence of other phases (e.g. the color superconducting phase)
was also suggested quite long ago\cite{Bar77}, this two-phase
structure became the standard picture of the QCD phase diagram for
more than two decades. In recent years, however, it has been
established that at low temperatures and medium densities several
other phases might appear (see e.g. Ref.\cite{rev} for recent
reviews). In particular, for the case of two light flavors it has
been shown that there should be a non-negligible region in the QCD
phase diagram where strongly interacting matter is a color
superconductor (2SC phase)\cite{ARW98}. Unfortunately, due to
difficulties in dealing with finite chemical potential, ab initio
calculations (as e.g. lattice QCD) are not yet able to provide a
detailed knowledge of the QCD phase diagram\cite{Karsch:2003jg}.
Thus, most theoretical approaches are based on the use of
effective models of QCD. Among them, the Nambu-Jona-Lasinio
model\cite{NJL61} is one of the most popular. In this model the
quark fields interact via local four point vertices which are
subject to chiral symmetry. If such interaction is strong enough,
chiral symmetry is spontaneously broken at zero temperature and
density, and pseudoscalar Goldstone bosons appear. It has been
shown by many authors that when the temperature and/or density
increase, the chiral symmetry is restored\cite{VW91}. When
effective quark-quark interactions are added to the effective
lagrangian other phases also appear\cite{Bub04}. As an improvement
on the local NJL model, some covariant nonlocal extensions have
been studied in the last few years\cite{Rip97}. Nonlocality arises
naturally in the context of several successful approaches to
low-energy quark dynamics as, for example, the instanton liquid
model\cite{SS98} and the Schwinger-Dyson resummation
techniques\cite{RW94}. It has been also argued that nonlocal
covariant extensions of the NJL model have several advantages over
the local scheme. Several studies\cite{BB95,BGR02,SDS04} have
shown that these nonlocal models provide a satisfactory
description of the hadron properties at zero temperature and
density. Recently\cite{GDS00}, the characteristics of the chiral
phase transition have been investigated within this kind of
models. The aim of the present work is to extend such studies to
the case in which a superconducting phase can appear.

This article is organized as follows. In Sec.\ II we introduce the
model and formalism. Our results for some specific nonlocal regulators
are presented in Sec.\ III. In Sec. IV we analyze the dependence of
our results on the strength of the diquark correlations. Finally,
in Sec.\ V we give our conclusions.

\section{The formalism}

Let us begin by stating the Euclidean action for the nonlocal
chiral quark model in the case of two light flavors and $SU(2)$
isospin symmetry. For the purpose of the present study it is
enough to only consider interactions in the scalar quark-antiquark
and quark-quark channels. Thus, one has
\begin{eqnarray}
S_E &=& \int d^4 x \left\{ \bar \psi (x) \left(- i \rlap/\partial  +
m_c \right) \psi (x) \right. \nonumber \\
& & \qquad \qquad \left. - \frac{G}{2}\ \left[ \bar q(x) \ q(x) \right]^2 -
\frac{H}{2}\
\left[ \bar q(x) \ i \gamma_5 \tau_2 \lambda_2 \ q_C(x) \right]
\left[ \bar q_C(x) \ i \gamma_5 \tau_2 \lambda_2 \ q(x) \right]
\right\} \ ,
\label{action}
\end{eqnarray}
where $m_c$ is the current quark mass and $\tau_2$ and $\lambda_2$ are Pauli and
Gell-Mann matrices corresponding to the flavor and color groups, respectively.
The delocalized quark fields $q(x)$ are defined in terms of the quark fields $\psi(x)$
as
\begin{equation}
q(x) = \int d^4 y \ r(x-y) \ \psi(y)
\end{equation}
where the function $r(x-y)$ is a nonlocal regulator. It can be translated into
momentum space,
\begin{equation}
r(x-z) = \int \frac{d^4p}{(2\pi)^4} \ e^{-i(x-z) p} \ r(p) \;.
\end{equation}
Lorentz invariance implies that $r(p)$ can only be a function of $p^2$. Hence we will use for
the Fourier transform of the regulator the form $r(p^2)$ from now on.
In addition, in Eq.(\ref{action}) we have used
\begin{eqnarray}
q_C(x) = \gamma_2\gamma_4 \ \bar q\ \!^t(x) \qquad ; \qquad
\bar q_C(x) = q\ \!^t(x)\ \gamma_2\gamma_4.
\end{eqnarray}

{}The partition function for the model is defined as
\begin{equation}
{\cal Z} = \int {\cal D} \bar\psi {\cal D} \psi \  e^{-S_E(\mu,T)} \ ,
\label{zcero}
\end{equation}
where the Euclidean action at finite temperature $T$ and chemical potential $\mu$
is obtained from Eq.~(\ref{action}) by going to momentum space and performing
the replacement
\begin{equation}
\int \frac{d^4 p}{(2\pi)^4}\; F(\vec p,p_4) \quad \to \quad  T \sum_{n=-\infty}^{\infty} \int \frac{d^3
\vec{p}}{(2\pi)^3}\; F(\vec p,\omega_n - i\mu)\;,
\end{equation}
where $\omega_n$ are the Matsubara frequencies corresponding to
fermionic modes, $\omega_n = (2 n+1) \pi T$. As in
Ref.\cite{GDS00} we are assuming here that the quark interactions
only depend on the temperature and chemical potential through the
argument of the regulators. To proceed it is convenient to perform
a standard bosonization of the theory. Thus, we  introduce the
sigma meson field $\sigma$ and the scalar diquark field $\Delta$
and integrate out the quark fields. In what follows we will work
within the mean field approximation (MFA), in which these bosonic
fields are expanded around their vacuum expectation values $\bar
\sigma$ and $\bar \Delta$ and the corresponding fluctuations
neglected. Within this approximation and employing the
Nambu-Gorkov formalism, the mean field thermodynamical potential
per unit volume reads
\begin{equation}
\Omega^\mf  =  - \frac{T}{V} \, \ln {\cal
Z}^\mf  = \frac{\bar \sigma^2}{2 G} + \frac{|\bar
\Delta|^2}{2 H} - T \sum_{n=-\infty}^{\infty} \int \frac{d^3
\vec{p}}{(2\pi)^3} \
\frac{1}{2} Tr \ln \left[ \frac{1}{T} \ S^{-1}(\bar \sigma,\bar \Delta) \right] \label{z}
\end{equation}
where $Tr$ stands for the trace over the Dirac, flavor, color and Nambu-Gorkov
bispinor indexes. The inverse propagator $S^{-1}(\bar \sigma,\bar \Delta)$ is
\begin{equation}
S^{-1}(\bar \sigma,\bar \Delta) = \left( \ba{cc}
- \vec p \cdot \vec \gamma - \left( \omega_n - i\mu \right) \gamma^4 + \Sigma_p &
i \gamma_5 \tau_2 \lambda_2 \Delta_p  \\
i \gamma_5 \tau_2 \lambda_2  \Delta_p^* &
- \vec p \cdot \vec \gamma - \left( \omega_n + i\mu \right)\gamma^4 + \Sigma_p\ \!\!\! ^* \ea \right) \ ,
\end{equation}
where
\begin{equation}
\Sigma_p  =  m_c + \bar \sigma \ r^2_p \qquad ; \qquad
\Delta_p =  \bar \Delta  \ |r_p|^2
\end{equation}
and $r_p \equiv r\left( \vec p\ ^2 + (\omega_n - i\mu)^2 \right)$.
After some straightforward algebra $\Omega^\mf$ can be more explicitly
expressed as
\begin{eqnarray}
\!\!\!\!\! \Omega^\mf  & = &
\frac{\bar \sigma^2}{2 G} + \frac{|\bar \Delta|^2}{2 H} - \nonumber \\
& & 2 \
T \sum_{n=-\infty}^{\infty} \int \frac{d^3
\vec{p}}{(2\pi)^3} \left\{ 2 \ln \left[ \frac{ \left(A_p +
|\Delta_p|^2\right)^2 - B_p - 4 C_p^2} {T^4} \right] +  \ln \left[
\frac{A_p^2 - B_p - 4 C_p^2}{T^4} \right] \right\} \label{omega}
\end{eqnarray}
where,
\begin{eqnarray}
A_p  =  \omega_n^2 + {\vec p}\ ^2 + \mu^2 + |\Sigma_p|^2 , \qquad
B_p  =  4 {\vec p}\ ^2 \left( \mu^2 + \mbox{Im}^2 \Sigma_p  \right)  , \qquad
C_p  =  \mu \ \mbox{Re} \Sigma_p +  \omega_n \ \mbox{Im} \Sigma_p\ .
\label{def1}
\end{eqnarray}
For finite values of the current quark mass, $\Omega^\mf$ turns out to be divergent.
The regularization procedure used here amounts to define
\begin{equation}
\Omega^\mf_{(reg)} =
\Omega^\mf - \Omega^{free}
+ \Omega^{free}_{(reg)},
\label{omegareg}
\end{equation}
where $\Omega^{free}_{(reg)}$ is the regularized expression for
the thermodynamical potential of a free fermion gas,
\begin{equation}
\Omega^{free}_{(reg)} = -12\ T \int \frac{d^3 \vec{p}}{(2\pi)^3}\;
\left[ \ln\left( 1 + e^{-\left( \sqrt{\vec{p}^2+m_c^2}-\mu
\right)/T} \right) + \ln\left( 1 + e^{-\left(
\sqrt{\vec{p}^2+m_c^2}+\mu \right)/T} \right) \right].
\label{freeomegareg}
\end{equation}
The mean field values $\bar \sigma$ and $\bar \Delta$ are obtained from
the coupled pair of gap equations
\begin{equation}
\frac{ d \Omega^\mf_{(reg)}}{d\bar \sigma} = 0
\qquad , \qquad
\frac{ d \Omega^\mf_{(reg)}}{d\bar \Delta} = 0 \ .
\end{equation}
The explicit form of these equations is
\begin{eqnarray}
0 &=& \bar \sigma - 16\ G\ T \sum_{n=-\infty}^{\infty} \int \frac{d^3
\vec{p}}{(2\pi)^3}  \left\{
\frac{(A_p + |\Delta_p|^2) E_p - F_p - 2 C_p G_p}
{\left(A_p + |\Delta_p|^2 \right)^2  - B_p - 4 \; C_p^2} +
\frac{A_p E_p - F_p - 2 C_p G_p }{2 \left( A_p^2  - B_p - 4 C_p^2 \right) } \right\}
\ , \nonumber\\
& & \nonumber \\
0 &=& |\bar \Delta| \left[ 1 - 16\ H\ T \sum_{n=-\infty}^{\infty} \int \frac{d^3
\vec{p}}{(2\pi)^3} \left\{|r_p|^4 \;
\frac{A_p + |\Delta_p|^2}{\left(A_p + |\Delta_p|^2 \right)^2  - B_p - 4 \; C_p^2}
\right\} \right],
\label{gapeq}
\end{eqnarray}
where, in addition to the definitions given in Eq.(\ref{def1}), we have
used
\begin{eqnarray}
E_p  =  \mbox{Re} \Sigma_p \; \mbox{Re} \; r^2_p + \mbox{Im}
\Sigma_p \; \mbox{Im} \; r^2_p\ ,
\qquad
F_p  =   2 \; {\vec p}\ ^2 \; \mbox{Im} \Sigma_p \; \mbox{Im} \; r^2_p \ , \qquad
G_p  = \mu \; \mbox{Re} \; r^2_p + \omega_n \; \mbox{Im} \; r^2_p\ .
\end{eqnarray}
It should be noticed that, in general, there might be regions for which there are more
than one solution for each value of $T$ and $\mu$. In such regions
we identify the stable solution by requiring it to be an overall minimum of the
potential.

Given the thermodynamic potential the expressions for all other relevant quantities
can be easily derived. For each flavor the quark-antiquark condensate
$\langle \bar \psi \psi \rangle$ and the quark density $\rho_q$ are given by
\begin{equation}
\langle \bar \psi \psi \rangle = \frac{ \partial \Omega^\mf_{(reg)}}{\partial m_c}
\qquad , \qquad
\rho_q = - \frac{ \partial \Omega^\mf_{(reg)}}{\partial \mu}\ .
\end{equation}
In the case of the quark-quark condensate an extra source term $
-\xi \ \bar{\psi}_C (x) i \gamma_5 \tau_2 \lambda_2 \psi(x)$ has
to be added to the effective action. It is easy to see that this
leads to a thermodynamic potential $\Omega_{MFA}(\xi)$ which has
the form given in Eq.(\ref{omega}) but where $\Delta_p$ has been
replaced by $\Delta_p - \xi$. Then, we get
\begin{equation}
\langle \psi \psi \rangle = \left. -\frac{ \partial
\Omega^\mf_{(reg)}(\xi)} {\partial \xi}\right|_{\xi=0}\ .
\end{equation}
Finally, a magnitud which is important to determine the characteristic of the
chiral phase transition is the chiral susceptibility $\chi$. It can be calculated as
\begin{equation}
\chi = - \frac{ \partial^2 \Omega^\mf_{(reg)}} {\partial m_c^2} =
- \frac{ \partial \langle \bar \psi \psi \rangle} {\partial m_c}\ .
\end{equation}

\section{Numerical results for different regulators}

In this section we concentrate on the numerical results obtained
for two different regulators often used in the literature: the
Gaussian regulator and the Lorentzian regulator. In each case we
have considered $G$, $m_c$ and $\Lambda$ as input parameters fixed
so as to reproduce the phenomenological values of the chiral
condensate, pion mass and pion decay constant at vanishing
temperature and densities\cite{BB95,GDS00}. Moreover, we have set
$H/G = 3/4$ as implied by, for example, OGE interactions
\cite{ARW98,Bub04}. The dependence of our results on this ratio
will be discussed in the following section.

\subsection{Gaussian regulator}

In this case the regulator is given by
\begin{equation}
r(p^2) = \exp(-p^2/2\Lambda^2),
\end{equation}
where $\Lambda$ plays the role of a cut-off parameter. We have
considered two different sets of parameters. Set I corresponds to
$G = 50\ GeV^{-2}$, $m_c = 10.5\ MeV$ and $\Lambda = 627\ MeV$
while Set II to $G = 30\ GeV^{-1}$, $m = 7.7\ MeV$ and $\Lambda =
760\ MeV$. Although for both sets the zero temperature and density
properties mentioned above are well reproduced, in the case of Set I
the quark propagator has no purely real poles while for Set II it
does. Thus, following Ref.\cite{BB95}, Set I might be interpreted
as a confining one, since quarks cannot materialize on-shell in
Minkowski space.

\begin{figure}[ht]
\begin{center}
\centerline{\psfig{figure=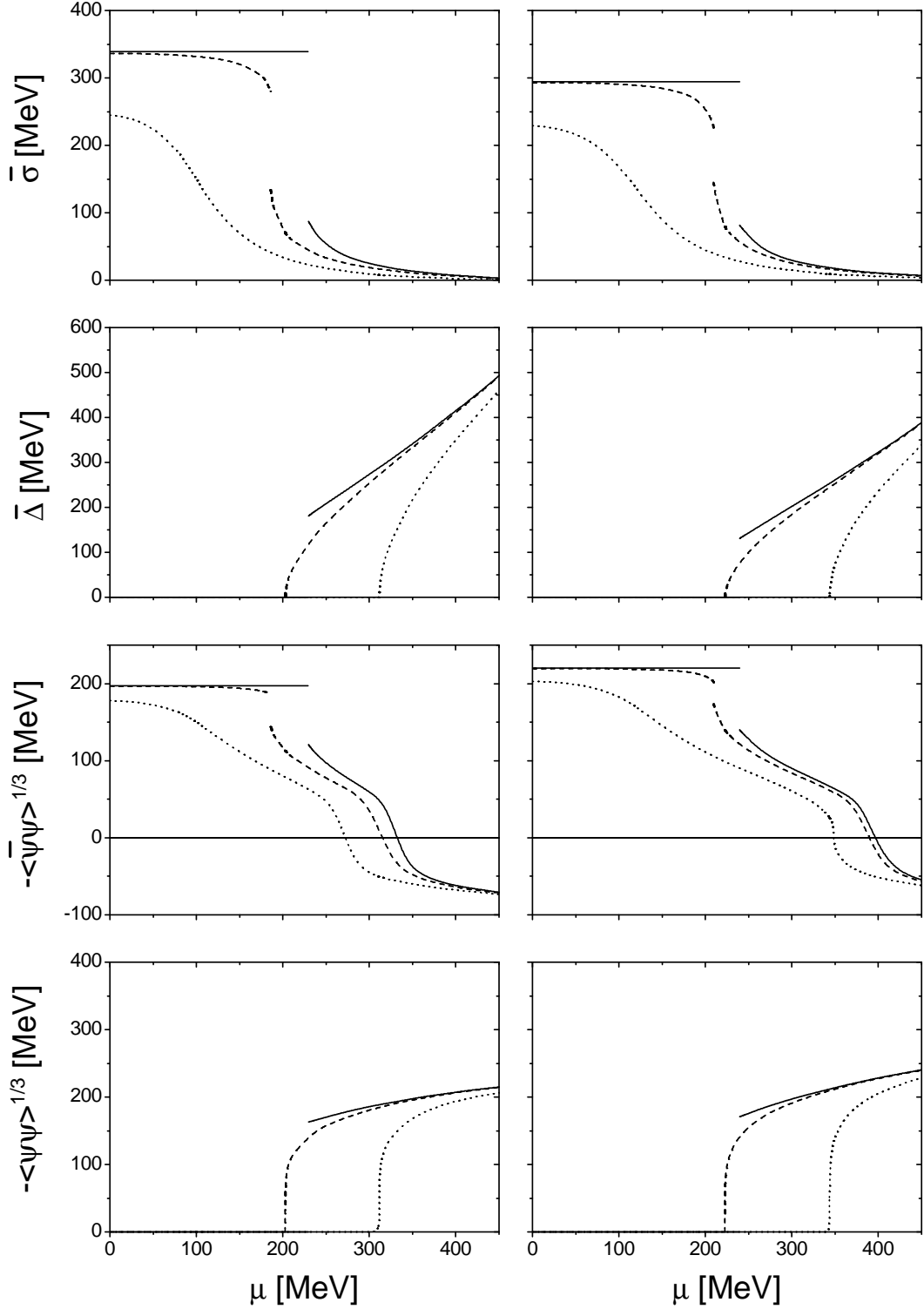,height=20cm}}
\caption{Behavior of mean fields $\bar{\sigma}$, $\bar{\Delta}$
and condensates $\langle {\bar{q}} q \rangle$, $\langle q q
\rangle$ for the Gaussian regulator, as a function of chemical
potential for three different values of the temperature.
Left panels correspond to Set I and right ones to Set II.
Full lines correspond to $T=0$, dashed lines to $T = 67 MeV$ for Set I
($T = 57 MeV$ for Set II) and dotted lines to $T = 100 MeV$.
\label{unos} }
\end{center}
\end{figure}

The mean field values $\bar \sigma$ and $\bar \Delta$ as a
function of $\mu$ and $T$ are obtained by numerically solving
Eqs.(\ref{gapeq}). The corresponding results as a function of
$\mu$ for various values of $T$ are displayed in Fig.\ref{unos},
where left panels correspond to Set I while right ones to Set II.
It can be seen that, for small values of $T$ and $\mu$ (full lines
in Fig.\ref{unos}), the system is in the chiral phase for which
$\bar{\sigma} \neq 0$ and $\bar{\Delta} = 0$. If we increase $\mu$
keeping $T$ fixed, at some critical value of $\mu$ there is sudden
drop of $\bar{\sigma}$ and a simultaneous sudden increase in
$\bar{\Delta}$ so that we get into the 2SC phase characterized by
$\bar{\sigma} \cong 0$ and $\bar{\Delta} \neq 0$. In particular,
the values of the diquark gap at the critical chemical potential
and $T=0$ can be found in Table 1. If we repeat the process with a
higher value of $T$ something similar happens until we reach the
``triple point" (3P). For temperatures slightly higher than
$T_{3P}$ (dashed lines in Fig.\ref{unos}) the sudden drop in
$\bar{\sigma}$ and the increase $\bar{\Delta}$ start to happen at
two different values of $\mu$. Between these values of $\mu$ we
have $\bar{\sigma} \cong 0$ and $\bar{\Delta} = 0$. Moreover there
is no discontinuity in the behavior of $\bar{\Delta}$ as a
function of $\mu$. For temperatures above the ``end point" (EP)
the discontinuity in $\bar{\sigma}$ also disappears (dotted lines
in Fig.\ref{unos}). Finally for temperatures above the critical
temperature for $\mu=0$, $T_c(\mu=0)$,  we get $\bar{\sigma} \cong
0$ for all values of $\mu$. In the region corresponding to the
crossover the transition point is defined by the point at which
the chiral susceptibility $\chi$ has a maximum. The positions of
the different critical points are summarized in Table 1. Also
shown in Fig.\ref{unos} are the corresponding quiral $\langle \bar
\psi \psi \rangle$ and diquark $\langle \psi \psi \rangle$
condensates. Their behavior is quite similar to those of the mean
field values $\bar \sigma$ and $\bar \Delta$, respectively. It is
worthwhile to mention that the fact that the chiral condensate
approaches some positive value for large values of $\mu$ is due to
the substraction scheme used to regularize the thermodynamical
potential, Eq.(\ref{omegareg}). In fact, it is not difficult to
see that for finite values of $m_c$ the regularized free
thermodynamical potential $\Omega^{free}_{(reg)}$,
Eq.(\ref{freeomegareg}), has such behavior.

\begin{figure}[ht]
\begin{center}
\centerline{\psfig{figure=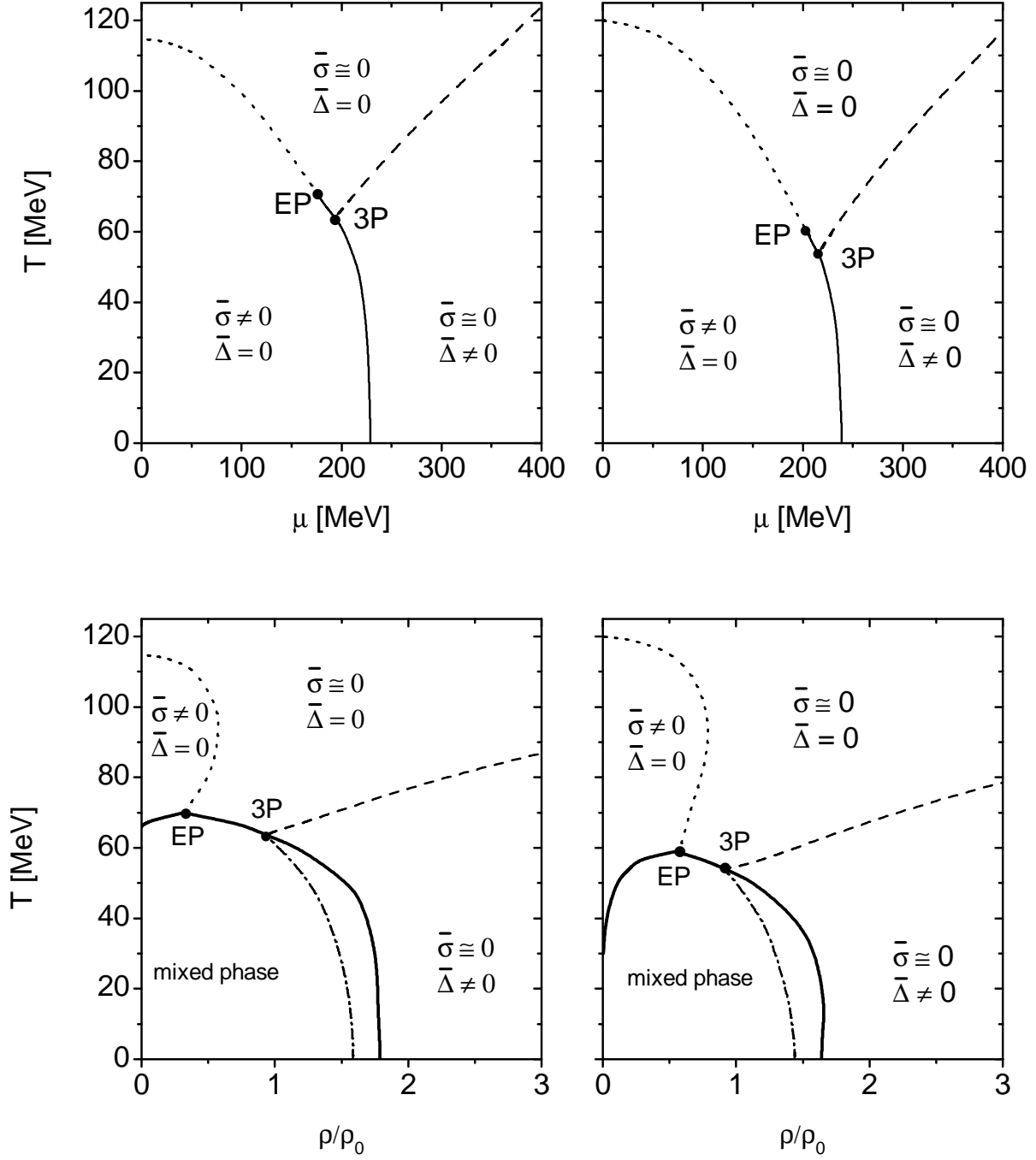,height=18cm}}
\caption{Phase diagrams for the Gaussian regulator.
Left panels correspond to Set I and right ones to Set II.
Upper panels display the $T-\mu$ phase diagrams and the
lower ones the $T-\rho/\rho_0$ phase diagrams. Full lines
indicate first order transition lines, dashed lines correspond
to second order transition lines and dotted lines to crossover
lines. The dash-dotted line in the lower panels indicates the section
of the transition line corresponding to $H=0$.
\label{dos} }
\end{center}
\end{figure}

The corresponding phase diagrams are displayed in Fig.\ref{dos}.
Again left panels correspond to Set I and right ones to Set II. On
the other hand, the upper panels correspond to the phase diagrams
in the $T-\mu$ plane and the lower ones to the diagrams in the
$T-\rho/\rho_0$ plane, where the nuclear matter density $\rho_0 =
1.3 \times 10^6 MeV$. In all the cases we have indicated with full
lines the first order transition lines, with dashed lines the
second order transition lines and with dotted lines the lines
corresponding to the crossover between the chiral phase and the
weakly interacting quark phase.

As it is clear from the $T-\mu$ phase diagrams, at the triple point
the three phases can coexist. It is interesting to remark
that for values of the chemical potential below $\mu_{3P}$ the
transition line (both the first order
and crossover sections) coincides exactly with that obtained in
the absence of diquark correlations ($H=0$). On the other hand,
for $\mu > \mu_{3P}$ the first order transition line is different
from that obtained in Ref.\cite{GDS00}. This is more clearly seen
in the corresponding $T-\rho/\rho_0$ diagrams. There we have
indicated with dash-dotted line the first order transition line
corresponding to $H=0$. As we observed the existence of diquark
correlations increases the size of the mixed phase. Note that for
$T < T_{3P}$ such mixed phase is composed by the chiral and 2SC
phases while for $T > T_{3P}$ it is a mixture of the chiral and
the free quark gas phases.

As for the comparison between the results of Set I and II, we see
that they are qualitatively very similar, with only small
quantitative differences in the position of the critical points.

\subsection{Lorentzian regulator}

\begin{figure}[ht]
\begin{center}
\centerline{\psfig{figure=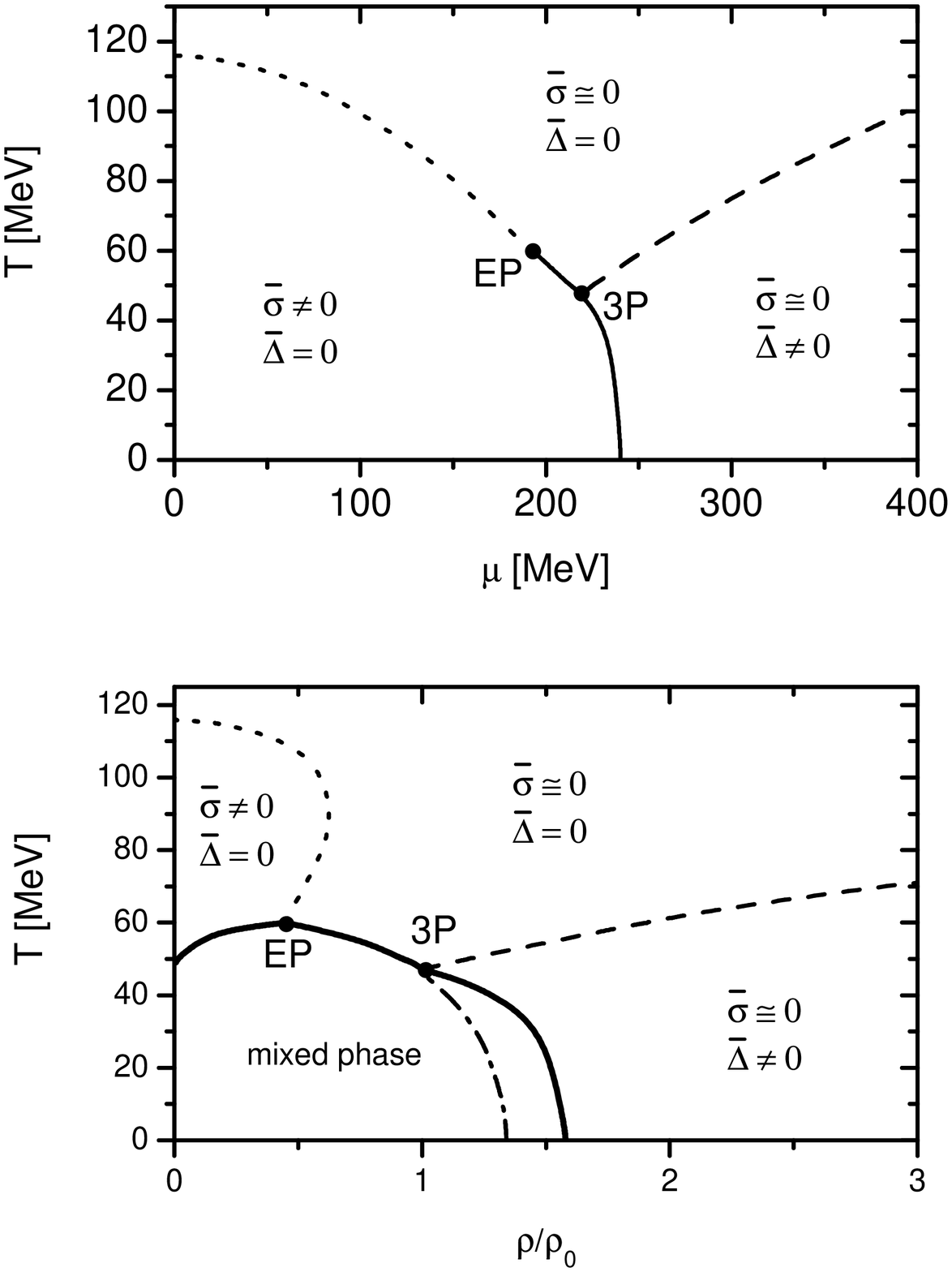,height=18cm}} \caption{Phase
diagrams for the Lorentzian regulator. Upper panels display the
$T-\mu$ phase diagrams and the lower ones the $T-\rho/\rho_0$
phase diagrams. Full lines indicate first order transition lines,
dashed lines correspond to second order transition lines and
dotted lines to crossover lines. The dash-dotted line in the lower
panels indicates the section of the transition line corresponding
to $H=0$. \label{tres} }
\end{center}
\end{figure}

The Lorentzian regulator we have considered has the form
\begin{equation}
r(p^2) = \frac{1}{1 + p^2/\Lambda^2}.
\end{equation}
The input parameters are $G = 28.38\ GeV^{-2}$, $m_c =
4.57\ MeV$, $\Lambda = 940\ MeV$. Once again, we have solved the
gap equations for different values of the temperature and chemical
potential. The qualitative behavior of the mean field values and
condensates is very similar to that found using the Gaussian
regulator (Fig.\ref{unos}) and, thus, will not be explicitly
shown. Nevertheless it is interesting to note that, as shown in
Table 1, in this case the $T=0$ diquark gap at the critical chemical potential
$\bar \Delta_c$ is somewhat smaller. The corresponding phase diagrams in the
$T-\mu$ and $T-\rho/\rho_0$ planes are displayed in Fig.\ref{tres}. As in the
previous cases, we observe the existence of an ``end point", at
which the first order chiral phase transition line becomes a
crossover line, and a triple point at which the three phases
co-exist. The position of these points, which are quite similar to
those of the Gaussian regulator Set II, are given in Table 1.

\section{Dependence of the phase diagrams on the diquark coupling
constant}

In the previous section we have assumed $H/G=3/4$ as favored by
various effective models of quark-quark interactions. However,
this value is subject to rather large uncertainties. In fact, so
far there is no strong phenomenological constrain on the ratio
$H/G$. Thus, it is worthwhile to explore the consequences of
varying $H/G$ in the range $0 < H/G \leq 1$. Larger values of this
ratio are quite unlikely to be realized in QCD and might lead to color
symmetry breaking in the vacuum. Since the results
obtained above for different regulators are qualitatively very
similar in what follows we will only consider the Gaussian
regulator with the set of parameters Set II.

\begin{figure}[ht]
\begin{center}
\centerline{\psfig{figure=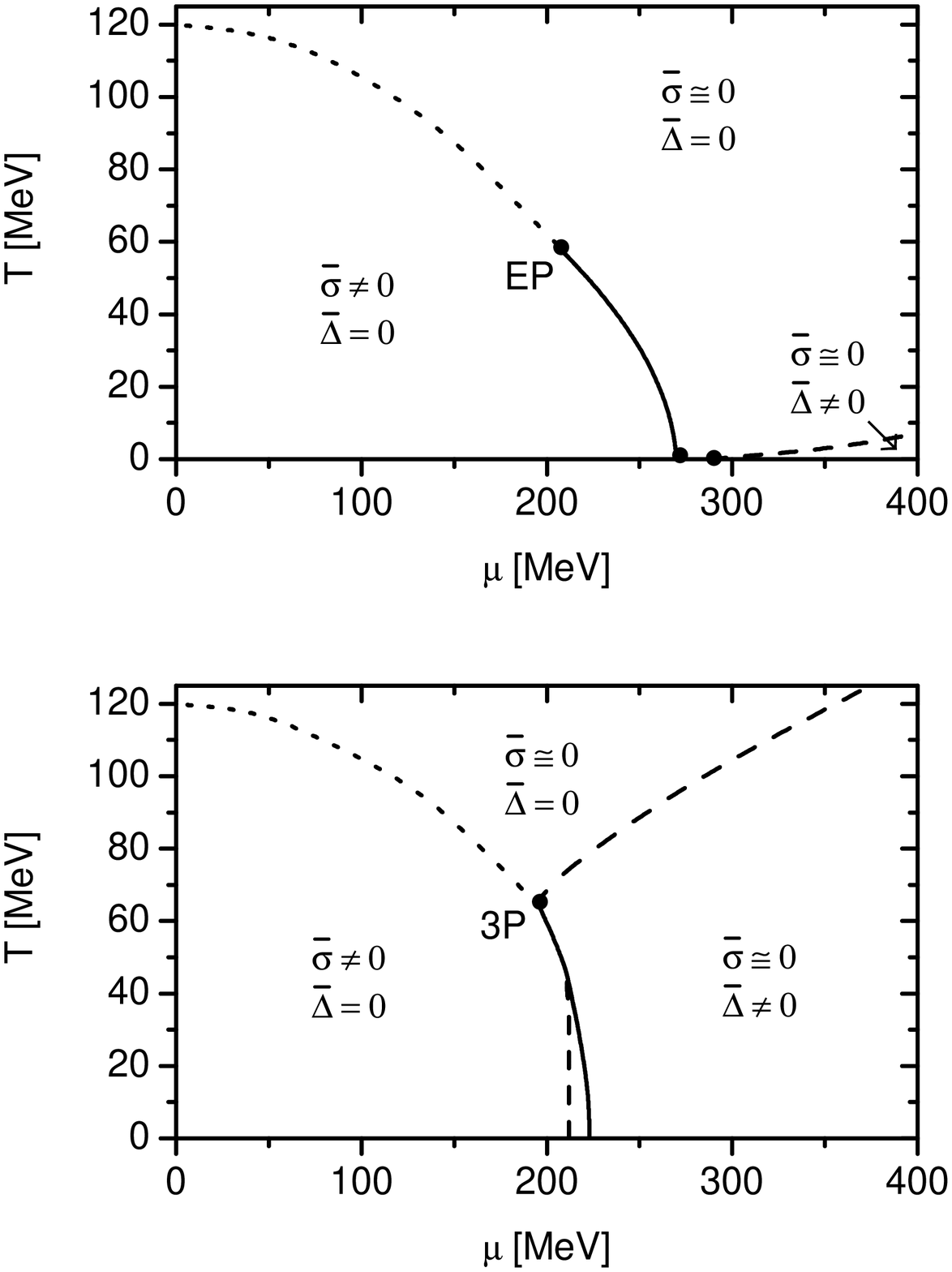,height=20cm}}
\caption{Phase diagrams for the Gaussian regulator (Set II) and
different values of the ratio $H/G$.
The upper panel corresponds to the ratio $H/G = 0.15$ while the lower
one to $H/G=0.9$.
\label{cinco} }
\end{center}
\end{figure}

For values of $H/G$ in the range $ 0.17 < H/G < 0.82 $ the
resulting phase diagrams look qualitatively similar to the one
displayed in Fig.\ref{dos}, although the details (in particular
the position of the critical points, see below) do depend on
$H/G$. For $H/G < 0.17$ there is a qualitative change since the
triple point ceases to exist. In this case, even at very low
temperatures, as we increase $\mu$ at some point we find a first
order phase transition between the chiral phase and the free quark
gas phase. For values slightly above this critical $\mu$ we have
$\bar \sigma \simeq \bar \Delta = 0$. If we continue to increase
$\mu$ we find a second order phase transition between the free
quark gas phase and the 2SC phase. The corresponding $T-\mu$ phase
diagram is shown in the upper panel of Fig.\ref{cinco}. Note that
in this phase diagram the transition line between the chiral and
the free quark gas phases (both its crossover and first order
sections) coincides exactly to the one obtained for $H=0$. For
$H/G > 0.82$ the situation is again qualitatively different, since
in such range the triple point and the end point merge together.
In fact, as $H/G$ comes closer from below to $H/G=0.82$ the first
order transition line that connects the 3P and the EP becomes
shorter and shorter and at this particular value it disappears.
Moreover, for $H/G > 0.82$ there is small region at low
temperatures at which both $\bar \sigma$ and $\bar \Delta$ take
non-vanishing values. This region is separated from the chiral
phase by a second order transition line and from the 2SC for a
first order transition line. This situation is illustrated in the
lower panel of Fig.\ref{cinco} where the $T-\mu$ phase diagram for
$H/G=0.90$ is shown. The possible existence of this type of phase
was already noticed in Refs.\cite{Rapp:1999qa,BVY03}.

\begin{figure}[ht]
\begin{center}
\centerline{\psfig{figure=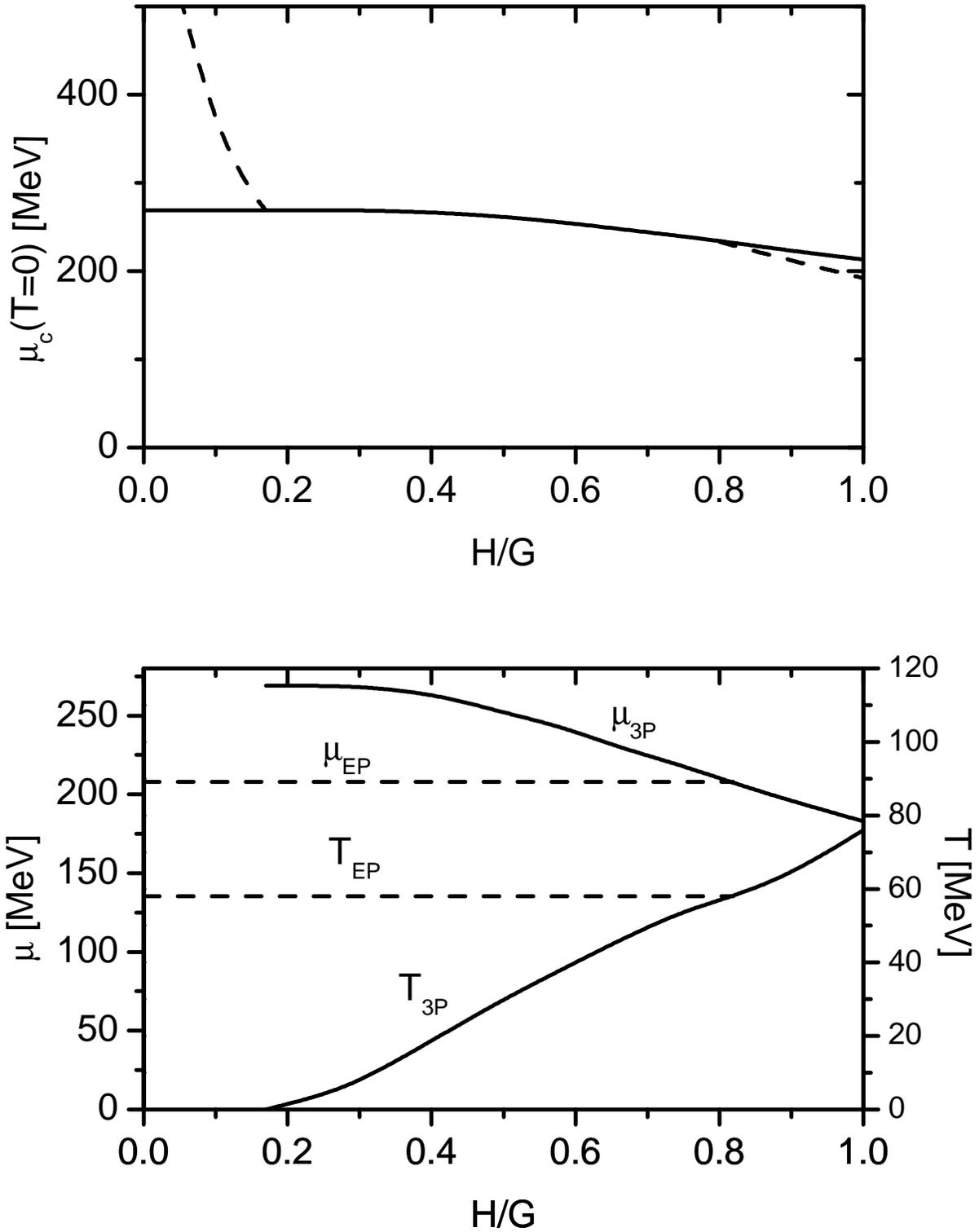,height=20cm}}
\caption{Behavior of the critical points for the Gaussian regulator (Set II)
as a function of $H/G$.
The upper panel displays the critical chemical potentials at $T=0$. The
lower panel shows the position of the triple and end points.
\label{seis} }
\end{center}
\end{figure}

The behavior of the critical points as function of $H/G$ is
displayed in Fig.\ref{seis}. In the upper panel we show the
position of critical chemical potential $\mu_c$ at $T=0$. The full
line indicates the first order $\mu_c$ while the dashed line the
second order one. In the range $ 0.17 < H/G < 0.82 $ we have only
a first order $\mu_c$. For values below such $\mu_c$ the system is
in the chiral phase while for values above is in the 2SC. For
values $H/G < 0.17$ we have that the second order $\mu_c$, i.e.
the point at which the second order transition line that separates
the 2SC and free quark gas phases meets the $\mu$-axis in the
$T-\mu$ phase diagram, grows rather fast as $H/G$ decreases,
signaling the almost disappearance of the 2SC phase for very small
values of the diquark coupling constant. For $H/G > 0.82$ a second
order $\mu_c$ appears again, but now {\it below} the first order
$\mu_c$. Thus, between these two critical chemical potentials we
have $\bar \sigma \neq 0$ {\it and} $\bar \Delta \neq 0$. In the
lower panel we display the position of the triple and end points
as functions of $H/G$. Note that the temperature scale is given to
the right while the chemical potential scale is given to the left.
For values of $H/G < 0.17$ only the the EP exists. Its positions
remains independent of the diquark coupling constant up to
$H/G=0.82$ where it meets the 3P, that appears at $H/G=0.19$ and
which position in temperature (chemical potential) increases
(decreases) as $H/G$ increases. For values of $H/G > 0.82$ both
critical points transform into a single one which position in
temperature (chemical potential) increases (decreases) as $H/G$
increases.

\section{Conclusions}

In this work we have studied the finite temperature and chemical
potential behavior of $SU(2)_f$ chiral quark models with nonlocal
covariant separable interactions in both the scalar
quark-antiquark and quark-quark channels. In our numerical
calculations we have considered two types of regulators: the
Gaussian regulator and the Lorentzian regulator. In all these
cases we have set the model parameters so as to reproduce the
empirical values of the pion mass and decay constant and to get a
chiral quark condensate in reasonable agreement to that determined
from lattice gauge theory or QCD sum rules. As for the ratio
between quark-quark and quark-antiquark interactions $H/G$ which
is not well constrained by phenomenology, we have in principle
adopted the standard value $H/G=3/4$ which follows from some
models of the QCD interactions. We find that in all cases the
phase diagram is quite similar. In particular, we obtain that for
two light flavors there are always two critical points: a ``triple
point" at which the second order transition line separating the
2SC and normal phases meets the first order transition line which
separate the chiral and 2SC phases at low temperatures; an ``end
point" which appears at higher temperatures and at which the first
order transition line becomes a crossover line. Of course, there
is also a critical temperature $T_c (\mu=0)$ above which the
chiral condensate always vanishes. As displayed in Table 1 the
values of $T_c (\mu=0)$ are in the range $115-120$ MeV, that is
somewhat below the values obtained in modern lattice simulations
which suggest $T_c(\mu =0) \approx 140 -
190$~MeV\cite{Karsch:2003jg}. With this in mind we note that our
predictions for the positions of the triple and end points are
very similar for all the cases considered. Perhaps, the only
noticeable difference between the different cases is the
prediction for the $T=0$ diquark gap at the critical chemical
potential, where we find values that range from 114 to 182 MeV.
These values are larger than those obtained within models with
instantaneous interactions \cite{rev}. It should be noticed (see
Fig.\ref{unos}) that in the present case the diquark gap also
registers a stronger increase with $\mu$ after the phase
transition. This leads to rather large values of the gap for
chemical potentials of the order of 400 MeV, above which strange
degrees of freedom have to be taken into account. Consequently,
for such values of $\mu$, we also get larger values (100-120 MeV)
for the critical temperature needed to go to the free quark gas
phase, as it can be easily seen comparing our phase diagrams with
those obtained in e.g. the NJL model\cite{Bub04,BVY03}.

In the final part of this work we have explored the consequences of
varying $H/G$ in the range $0 < H/G \leq 1$. Given the similarity
of the results obtained for the two regulators mentioned above we have considered
here only the Gaussian regulator with the set of parameters Set II. We found that for $H/G < 0.17$
there is no triple point. On the other hand, for $H/G > 0.82$ the triple and end point
merge and a phase where both the chiral and diquark condensates are non-negligible appears.
It is interesting to remark that the value $H/G=0.82$ is quite close to
the standard one $H/G=0.75$ used in most model calculations. Thus, it
would be important to sort out possible phenomenological consequences of
having quark matter with a phase diagram in which the 3P and EP coincide.

In this work we have neglected the strangeness degrees of freedom. To go
beyond the values of chemical potential considered here and, thus,
study for example the Color Flavor Locked phase in the context of models with non
local interactions their effect have to be included. Work in this direction is in progress.

\section*{Acknowledgements}

We thank D. Blaschke and  D. G\'omez Dumm for useful discussions. This work has
been partially supported by CONICET and ANPCyT under grants PIP
02368 and PICT 00-03-08580, respectively.

\vspace{3cm}
\begin{table}[b]
\begin{center}
\begin{tabular}{ccccccc}
\hline\hline
 & & & & & & \\
Regulator & \multicolumn{2}{c}{Triple Point} &
\multicolumn{2}{c}{End Point}
& \multicolumn{1}{c}{$T_c(\mu = 0)$} & \multicolumn{1}{c}{$\bar{\Delta}_c(T = 0)$}\\
 & $T_{3P}$ & $\mu_{3P}$
 & $T_{EP}$ & $\mu_{EP}$ &  &
\\ \hline Gaussian - Set I & 64 &193 &69 &180 &115 &182\\
Gaussian - Set II &54 &215 &58 &207 &120 &132\\
Lorentzian &48  &217 &59 &195 &116 &114\\
\hline\hline
\vspace{.5cm}
\end{tabular}
\caption{Critical temperatures, chemical potentials and
$\bar{\Delta}_c(T = 0)$ (all in MeV) for different regulators.}
\vspace{1cm}
\end{center}
\end{table}

\pagebreak

\end{document}